\documentclass[aps,pra,epsf,superscriptaddress,amsmath,amssymb,amsfonts,twocolumn,showpacs,nofootinbib]{revtex4-1}
\usepackage{amsmath}
\usepackage{subfigure}
\usepackage{graphicx}
\usepackage{dcolumn}
\usepackage{bm}
\usepackage[normalem]{ulem}
\usepackage{amssymb}
\usepackage{soul}
\usepackage{xcolor}
\usepackage{xparse}
\usepackage{color}
\usepackage{enumitem}
\usepackage{enumitem}
\usepackage{pgfplots}
\pgfplotsset{compat=1.9}
\usepackage{tikz}
\usepackage{hyperref}

\usepackage[normalem]{ulem}

\begin{document}

\preprint{APS/123-QED}


\title{Rogue waves in extended Gross-Pitaevskii
Models with a Lee-Huang-Yang correction}

\author{Sathyanarayanan Chandramouli}
\email{sathyanaraya@umass.edu}
\affiliation{Department of Mathematics and Statistics, University of Massachusetts Amherst, Amherst, MA 01003-4515, USA}

\author{S. I. Mistakidis}
\affiliation{Department of Physics and LAMOR, Missouri University of Science and Technology, Rolla, MO 65409, USA}

\author{G. C. Katsimiga}
\affiliation{Department of Physics and LAMOR, Missouri University of Science and Technology, Rolla, MO 65409, USA}%

\author{D. J. Ratliff}
\affiliation{Department of Mathematics, Physics and Electrical Engineering, Northumbria University, Newcastle, United Kingdom}%

\author{D. J. Frantzeskakis}
\affiliation{Department of Physics, National and Kapodistrian University of Athens, Athens, Greece}%

\author{P. G. Kevrekidis}
\affiliation{Department of Mathematics and Statistics, University of Massachusetts Amherst, Amherst, MA 01003-4515, USA}%

\affiliation{Department of Physics, University of Massachusetts Amherst, Amherst, MA 01003, USA}

\affiliation{Theoretical Sciences Visiting Program, Okinawa Institute of Science and Technology Graduate University, Onna, 904-0495, Japan}

\date{\today}

\begin{abstract}

We explore the existence and dynamical generation of rogue waves (RWs) within a one-dimensional quantum droplet-bearing environment. RWs are computed by deploying a space-time fixed point scheme to the relevant extended Gross-Pitaevskii equation (eGPE). 
Parametric regions where the ensuing RWs are different from their counterparts in the nonlinear Schr\"odinger equation are identified.   
To corroborate the controllable generation, relevant to ultracold atom experiments, of these rogue patterns we exploit two different protocols. The first is based on interfering dam break flows emanating from Riemann initial conditions and the second refers to the gradient catastrophe of a spatially localized waveform. 
A multitude of possible RWs are found 
in this system, spanning waveforms reminiscent of the Peregrine soliton, its spatially periodic variants—namely, the Akhmediev breathers—and other higher-order RW  solutions of the nonlinear Schrödinger equation. 
Key elements of the shape of the corresponding 
eGPE RWs traced back to non-integrability and the presence of competing interactions are discussed.  
Our results set the stage for probing a multitude of unexplored rogue-like waveforms in such  mixtures with competing interactions and should be accessible to current ultracold atom experiments. 
\end{abstract}

\maketitle

\section{\label{sec:introduction}INTRODUCTION} 
 A highly intriguing class of
 waveforms of emerging interest in  
 physical settings featuring a degree
 of self-focusing (attractive) nonlinear
 interactions is the family of {rogue waves} ({RWs}). 
Typically, these are elevation waves with amplitudes several times larger than their background, with a high degree of spatio-temporal localization which is perceived as 
their ``appearance out of nowhere and disappearance without a
trace''~\cite{akhmediev2009waves,kharif2003physical}. They are of wide interest in various fields ---where they have been {\it experimentally} observed---  such as hydrodynamics and water waves~\cite{chabchoub2011rogue,McAllister2019,chabchoub2012experimental,chabchoub2012observation,chabchoub2012observation2,chabchoub2014time,chabchoub2012super,slunyaev2023rogue} 
(including also capillary waves~\cite{Shats_RWs}), nonlinear optics~\cite{nat_solli,Kibler2010,Kibler2012,Dudley2014,Frisquet2016,tlidi2022rogue},
plasmas~\cite{Bailung2011,Sabry2012,Tolba2015}, superfluid  helium~\cite{ganshin2008observation}, ultracold atomic Bose-Einstein condensates (BECs)~\cite{romero2024experimental,mossman2024nonlinear}, and so on.  

The cubic nonlinear Schr{\"o}dinger (NLS) equation (\emph{alias} Gross-Pitaevskii equation (GPE) in free space) is a key model for understanding the properties of RWs. Specifically, there  appears to be a strong connection between RWs and the rational solutions of the NLS 
\cite{akhmediev2009waves,dysthetrulsen,shrira}, and particularly the Peregrine soliton (PS) \cite{peregrineanziam}. The latter, is a spatio-temporally localized rational solution of the focusing NLS 
of peak amplitude
roughly three times above that of the plane wave background, which represents a prototype 
among nonlinear RW structures. PSs have been observed in various
contexts, including nonlinear optics \cite{Kibler2010}, multi-component plasmas \cite{bailung2011observation},  fluid mechanics \cite{chabchoub2011rogue}, and ultracold 
atomic BECs~\cite{romero2024experimental}. A hierarchy of higher-order rational solutions of the cubic GPE have also been studied as potential candidates for RWs, both theoretically \cite{akhmediev2009rogue,tan2022super}
---also in BECs~\cite{adriazola2025experimentallytractablegenerationhighorder}--- and experimentally \cite{chabchoub2012observation,chabchoub2012super}. Mechanisms that have been suggested for the formation of RWs and PSs, such as the modulational instability (MI) and gradient 
catastrophe, are associated with the focusing 
(i.e., involving attractive interatomic interactions
in BECs)
nature of the nonlinearity; see, e.g., Refs.~\cite{dysthe2008oceanic,bertola2013universality}.

An alternative perspective of the PS is that of an infinite period  limiting
solution of the spatially (temporally) periodic Akhmediev~\cite{akhmediev2009waves} (Kuznetsov-Ma~\cite{kuznetsov1977solitons}) breather. 
Generalized variants of such structures have also appeared recently which include the so-called ``cnoidal RWs", i.e., localized lumps of large amplitude mounted atop a periodic wave background envelope \cite{kedziora2014rogue,DEP1,ChenPelinovsky2024}. 
RW excitations have also been studied in integrable generalizations of the cubic NLS   \cite{wang2013breather,yang2015rogue,ankiewicz2014extended} but also in various nonintegrable versions thereof 
in a perturbative \cite{ward2019evaluating,ward2020rogue,ankiewicz2014extended} and non-perturbative context \cite{zakharov2010shape,calini2008rogue}. Given that the vast majority of physically
realistic models are non-integrable, there
is a pressing need to explore methodologies
for understanding, characterizing and
 predicting RWs in such 
non-integrable systems.

Currently, several mechanisms 
are known to lead to RW generation including (i) localized perturbations on top of
modulationally unstable backgrounds that evolve into these ``extreme'' (or ``freak'') wave phenomena~\cite{ablowitz1990homoclinic}, (ii) the gradient catastrophe mechanism~\cite{bertola2013universality} and iii) interfering dam break flows (DBFs) \cite{el2016dam}. Gradient catastrophe is associated with the self-focusing dynamics of a wavepacket, which in the integrable NLS limit, leads to a space-time pattern whose features differ in various subregions divided by nonlinear caustics \cite{bertola2013universality}, across which a RW lattice can be seen. DBFs, on the other hand, represent a key feature in one-dimensional (1D) attractive environments, whose generation is the consequence of the universal stage of MI \cite{biondini2016universal,gurevich1993modulational}. They arise from the celebrated Riemann problem \cite{el2016dam,gurevich1993modulational}, and thus can be viewed as shock waves, which have been observed both in nonlinear optics \cite{wan2010diffraction} and BECs \cite{kh2005dynamics,mossman2024nonlinear}. Interestingly, the interference of such DBFs (referred to as a \textit{box} problem) is known to generate 
RW lattices of increasing \textit{genera} in the case of the integrable NLS models  \cite{el2016dam}. Although the above mentioned mechanisms
have been systematically analyzed in the 
integrable limit, far less is known
about the extent of their applicability
under different types of non-integrable
perturbations. Nevertheless, similarly, e.g.,
to the nonlinear stage of MI (which is central to the 
existence of RWs)~\cite{biondini2018universal},
it is anticipated that they will persist
in some form in non-integrable media.

On the other hand, recent experimental advances in the ultracold realm revealed another phase of matter, the so-called quantum droplet~\cite{luo2021new,mistakidis2023few,khan2022quantum}.   Such many-body self-bound states have been observed in both homonuclear~\cite{cheiney2018bright,semeghini2018self,cabrera2018quantum} and heteronuclear~\cite{d2019observation,burchianti2020dual} bosonic mixtures. They exist due to the competition between mean-field repulsion (attraction) and quantum fluctuation attraction (repulsion) in 1D~\cite{petrov2016ultradilute} (3D~\cite{petrov2015quantum}) and are commonly modeled by an appropriate extended GPE (eGPE) framework. The latter, incorporates the Lee-Huang-Yang (LHY) correction, 
accounting for the first-order correction to the mean-field energy functional.  
A plethora of coherent structures can be hosted in this eGPE setting, including bubbles~\cite{katsimiga2023interactions,katsimiga2023solitary,edmonds2023dark}, kinks~\cite{kartashov2022spinor,katsimiga2023interactions}, dark solitons~\cite{katsimiga2023solitary,saqlain2023dragging,gangwar2022dynamics}, 
traveling waves~\cite{paredes2025fatetravellingwavesboundary}
vortices and clusters thereof~\cite{li2018two,Bougas_vortex_drops,Michinel1}, as well as dispersive shock waves (DSWs)~\cite{katsimiga2023interactions,chandramouli2024dispersive}.  

However, despite the recent interest in quantum droplets,  neither the existence nor
the dynamical generation of RWs in such non-integrable eGPE setups has been demonstrated thus far, to the best of
our knowledge {despite the preliminary works in~\cite{lv2023excitation,lv2024excitation}.} 
The scope of the present work is to fill
this gap by revealing the potential 
presence of RW configurations in 1D droplet-bearing eGPE environments featuring competing attractive-repulsive interactions. 
Indeed, it is important to bear in mind
that such findings contribute to a deeper
understanding of RWs and their potential
existence in the wider setting of 
non-integrable media with competing
interactions. These, in turn, are expected
to lead to extensions in other fields
such as nonlinear optics,  where such
nonlinear competition is 
rather widespread, e.g., in the form of cubic-quintic nonlinearities~\cite{michinel3,araujo2,michinel2,araujo,BougasPRL}.
We anticipate lower but also higher-order RWs (HORWs) to form in our setting since droplet environments have been known for some time
to be prone to MI~\cite{mithun2020modulational}, given that there exist well-delineated parametric regions where the attractive LHY contribution dominates~\cite{chandramouli2024dispersive}. To address these questions and unveil the unprecedented emergence of RWs in droplet-bearing environments we employ the suitable 1D eGPE~\cite{petrov2016ultradilute} model which describes a symmetric (equal mass, atom number and repulsion per component) ultracold bosonic mixture experiencing  attractive intercomponent interactions.

In particular, we demonstrate the existence of two RW families pertaining to the eGPE model using a space-time fixed point scheme with doubly periodic basis representation.~\cite{ward2019evaluating}. Importantly, it is shown that there exist chemical potential regions where the identified RWs are similar to the standard PS of the cubic NLS \cite{peregrine1983water}, while for other chemical potentials the computed RWs are significantly different from their NLS counterparts.  
An \textit{effective} GPE with cubic nonlinearity is also constructed (i.e., deduced) from the eGPE model. The relevant reduction, in turn, facilitates the identification of RWs in the eGPE model, using as a building block for
the relevant construction the PS of the 
corresponding regular NLS model.

Subsequently, in order to facilitate 
the potential experimental observation of such entities, we propose two different dynamical protocols for generating RWs. These correspond to interfering DBFs, initiated by two symmetric (in space) Riemann-problems, as well as the gradient catastrophe scheme,  realized through a suitable spatially localized initial condition. Both of these initial conditions provide 
viable avenues for such a realization and 
can be crafted in ongoing ultracold atom experiments. Specifically, it is found that both protocols result in the dynamical generation of a plethora of distinct RW structures supported by the eGPE. These range from the PS of the NLS, to second-order RWs, spatially modulated RW lattices resembling Akhmediev-breathers of the NLS \cite{chabchoub2014hydrodynamics}, and persistent spatially localized solutions at long times. 
To validate our arguments on the existence of specific RW types we compare some computed eGPE solutions to integrable NLS PS and its HORW  generalizations. Differences between the shape of the eGPE RWs and those of the NLS  are discussed and are principally attributed 
to non-integrability and the presence of competing interactions. 

Our presentation is organized as follows. In Section~\ref{setup} we introduce the eGPE model under consideration and its NLS reduction used to gain insights on the observed RW solutions. Section~\ref{Results} discusses the existence and characteristics of RW families in the eGPE, while  Sec.~\ref{dynamics} describes the dynamical formation of RWs through the lens of counterpropagating DBFs and the gradient catastrophe. We conclude and offer a number of extensions of our results in Sec.~\ref{conclusions}. Appendix~\ref{appendix} compares DBFs predicted within the GPE and eGPE models.

\section{Extended model $\&$ its reduction}\label{setup}
The system under investigation consists of a 1D homonuclear bosonic 
mixture. Here, atoms of, e.g., $^{39}$K reside in two different hyperfine states, according to the corresponding 3D setup of~\cite{semeghini2018self} in free space. In the 1D case, the corresponding geometry is achieved by utilizing tightly confined transversal directions as compared to the elongated unconfined  longitudinal direction. Both states are assumed to share the same atom number ($N_1=N_2\equiv N$), feature equal intra-component repulsion, $g_{11}=g_{22}\equiv g>0$, and inter-component attraction, $g_{12}<0$, yielding a 1D droplet-bearing environment for $\delta g=g_{12}+g>0$.  
Under these assumptions, the two components become equivalent, allowing the mixture to be described by a reduced single-component 1D eGPE~\cite{petrov2015quantum,petrov2016ultradilute}, incorporating the leading-order LHY quantum correction. {Notably, the latter is attractive in 1D~\cite{petrov2016ultradilute} in sharp contrast to its repulsive 3D counterpart~\cite{petrov2015quantum}.} In dimensionless form, this model reads:
\begin{equation}
\label{GP-eqn}
i\psi_t+\frac{1}{2}\psi_{xx}
+f(\rho)
\psi=0,
\end{equation}
where $\psi(x,t)$ denotes the 1D  wavefunction, $\rho(x,t)=|\psi|^2$ is the local density, and $f(\rho)=-\rho+\sqrt{\rho}$ contains the nonlinear interaction terms. The quadratic nonlinearity captures the effectively 
attractive, 1D contribution of the LHY  correction, while the cubic term represents the standard mean-field repulsion. Finally, in Eq.~(\ref{GP-eqn}), energy is measured in units of $\hbar ^{2}/(m\xi ^{2})$, with $\xi =\pi \hbar ^{2}\sqrt{|\delta g|}/(mg\sqrt{2g})$ denoting the healing length and $m$  
is the atom mass. Also, time, length, and wavefunction are measured in units of $\hbar/\left(m\xi ^{2}\right) $, $\xi $ and $(2\sqrt{g})^{3/2}/(\pi \xi (2|\delta g|)^{3/4})$  respectively. {Representative dimensionless  evolution times considered herein correspond to $t\sim 10^3$, which refer to $\sim 245 $ms in dimensional units,  e.g., for a transverse trap frequency of $ \sim  2 \pi \times 650$Hz proximal to the droplet experiment of Ref.~\cite{cabrera2018quantum}. Recall
that we assume no longitudinal trap
herein.}

The competition between quadratic and cubic nonlinearities gives rise to distinct dynamical response regimes where the mean-field (LHY) repulsion (attraction) dominates
for densities $\rho_0\geq 0.25$ ($\rho_0<0.25$) as was argued in Ref.~\cite{chandramouli2024dispersive}. This density condition was also referred to as the hyperbolic-to-elliptic threshold therein. By considering small-amplitude perturbations on a homogeneous background with density $\rho_0\equiv |\psi_0|^2$ and Taylor expanding  $f(\rho)$ we obtain from Eq.~(\ref{GP-eqn}) the following attractive cubic GPE, governing the dynamics of {\it small amplitude disturbances},
\begin{equation}
\label{cubic-GP}
i\psi_t+\frac{1}{2}\psi_{xx}+\left(f(\rho_0)-\rho_0f'(\rho_0)\right)\psi+f'(\rho_0)|\psi|^2\psi=0,
\end{equation}
provided that $f'(\rho_0)>0$. 
Indeed, the vanishing of $f'(\rho_0)$ signals
the transition from self-focusing 
(for $\rho_0< 0.25$) to self-defocusing 
(for $\rho_0>0.25$) dynamics, in line with
the comments above. Notice also that the
third term in Eq.~(\ref{cubic-GP}) which is
linear in $\psi$ can
always be removed through a rescaling of
frequency via a gauge transformation~\cite{Sulem}. Motivated by the small amplitude approximation through the cubic GPE  model, we next explore the existence of RW-like excitations within the LHY dominated eGPE setting. 

\section{Existence of spatiotemporal RWs}
\label{Results}
To address the question of 
RW existence in the eGPE framework, we aim to compute families of space-time (periodic) solutions, as approximations to the space-time localized RW limit. These solutions are described by $\psi(x,t)=\phi(x,t) \exp(-i\mu_0 t)$, where $\mu_0{\in \mathbb{R}}$ denotes the chemical potential and $\phi(x,t){\in \mathbb{C}}$ is a spatio-temporal wavefunction satisfying the partial differential equation emanating from Eq.~(\ref{GP-eqn}): 
\begin{equation}
\label{eGPE-breather}
  {\bf L}_0[\phi]=  \mu_0 \phi+i\phi_t+\frac{1}{2}\phi_{xx}-|\phi|^2\phi+|\phi|\phi=0.
\end{equation}
Unlike the integrable cubic  
NLS (or GPE), the eGPE does not —to the best of our knowledge— admit explicit analytical formulas for rational solutions or space-time periodic RW-like generalizations thereof. 
Therefore, alternative numerical computation-based methods are necessary to establish their existence.

For this reason, here, we numerically compute space-time periodic RW solutions of Eq.~\eqref{eGPE-breather}  parameterized by $\mu_0$. 
This is accomplished by using a  doubly periodic spatiotemporal basis~\cite{ward2019evaluating,ward2020rogue} within a fixed-point iteration scheme based on the Newton–conjugate-gradient method \cite{yang2009newton} as it was previously proposed for non-integrable systems in Ref.~\cite{ward2019evaluating}. 
It is important to keep in mind that the solutions obtained herein
are actually doubly periodic (in space and
time) waveforms. Nevertheless, as was
observed in~\cite{ward2019evaluating},
for suitably wide choices of the space- and time-domains, they provide very good approximations
to the RW solutions in the case of the NLS
model where a comparison to the analytical
waveforms is accessible.

Unless stated otherwise, most computations in this work are carried out on spatial and temporal domains of size $L=30$ and $T=190$, respectively. These are  chosen to be sufficiently large to allow the system to relax to a {homogeneous background~\footnote{All of our converged numerical solutions exhibit a low residual, with an $\mathbb{L}^\infty$-norm (${\rm max}_{x,t}{\bf L}_0[\phi]$) less than $10^{-8}$.}. A critical aspect of this computation is the choice of appropriate space-time initial guesses within a suitable parameter regime, supplied by the PSs of Eq.~\eqref{cubic-GP}. %
Specifically, the PS waveform reads
\begin{eqnarray}
\psi^{(1)}(x,t)=\sqrt{\rho_0} e^{-i\mu_0 \tilde{t}}\left(1-\frac{4(1+2i\alpha_0 \tilde{t})}{1+4(\sqrt{\alpha_0}x)^2+4(\alpha_0 \tilde{t})^2}\right),
    \nonumber \\
\label{NLS-approximation-rogue}
\end{eqnarray}
where $\tilde{t}=t-\frac{T}{2}$ and $\alpha_0=\rho_0f'(\rho_0)=-\rho_0+\sqrt{\rho_0}/2$ is the dilatation factor.
Here, we emphasize that although Eq.~\eqref{NLS-approximation-rogue} is not an asymptotically valid approximation of the eGPE RW, it nevertheless remains sufficiently accurate for small $|\mu_0|\ll 1$  serving as a useful approximation or as an initial guess for seeding the fixed-point computations of the boundary value problem. {Thus, in the small $|\mu_0|$ limit, the eGPE RW is anticipated to resemble its integrable counterpart known from the cubic NLS \cite{el2016dam,bertola2013universality}, as we argue below. This fact partially highlights the robustness of NLS RWs in non-integrable systems and reflects the
generic nature of the relevant mechanisms
(going beyond the specific case example of
integrable systems).}

Figure~\ref{fig:1}(a) displays the space-time profile of a \textit{representative} eGPE waveform computed via the fixed-point scheme, corresponding to chemical potential $\mu_0 \approx -0.12$. Aside from mild background modulations, these solutions closely resemble a RW  behavior. Their peak profiles at $t = T/2$ are compared with the analytical PS solution of the reduced cubic GPE [Eq.~\eqref{cubic-GP}] in the low-density regime $\rho_0<0.02$ associated with $\mu_0>-0.1214$. In Fig.~\ref{fig:1}(c) a snapshot at $\tilde{t}=0$ of the numerically obtained solution is directly compared to the waveform of  
Eq.~\eqref{NLS-approximation-rogue} showing reasonable agreement in the wave core, despite the small 
differences in width, emanating from the
different nature of the nonlinearities in the
two models.
\begin{figure}
\centering    \includegraphics[width=\linewidth]{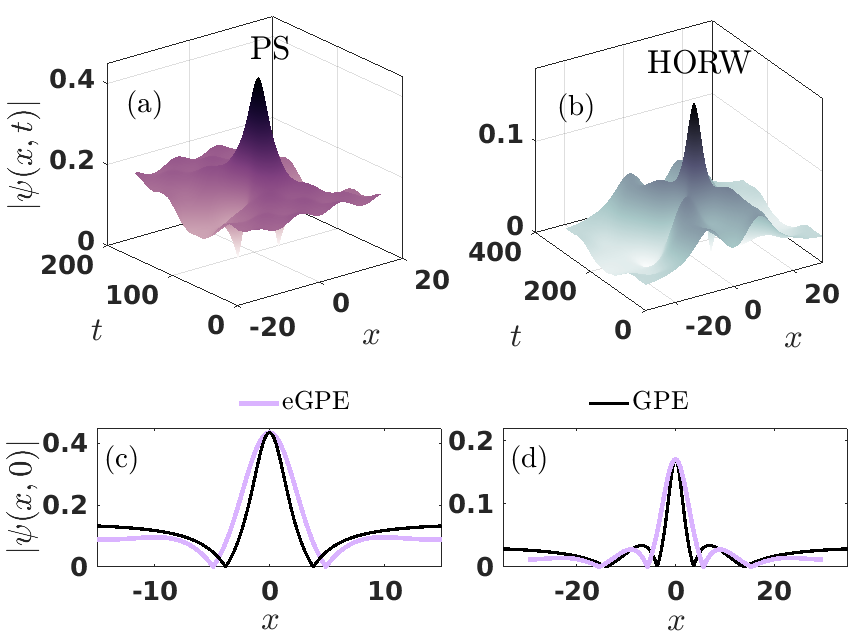}
\caption{
Representative eGPE RW moduli profiles, $|\psi(x,t)|$. 
(a) Space-time profile of a rogue waveform for $\mu_0\approx -0.12$, and (b) a second-order rogue waveform characterized by $\mu_0\approx-0.02$. (c) [(d)] Snapshot of the eGPE first-order [second-order] rogue waveform in (a) [(b)] at $\tilde{t}=0$ compared to a cubic GPE PS [second-order RW] possessing the same peak amplitude. 
Parameters used for the PS [HORW] are $L=30$ and $T=190$ [$L=60$ and $T=290$].
} \label{fig:1}
\end{figure}

Figure~\ref{fig:1}(b) illustrates the numerically identified spatiotemporal solution modulus, $|\psi(x,t)|$, of a HORW to the eGPE model. This higher-order entity features two nodes on either side of the central peak and it is built also atop a modulated background.
It turns out that this structure closely resembles a 2nd-order rational solution of the cubic GPE [see Eq.~\eqref{NLS-second-order}].
The latter, can be evidenced by directly comparing the profile of the numerically extracted solution at $\tilde{t}=0$ to the analytical expression available for the cubic GPE's HORW in the form~\cite{akhmediev2009rogue}
\begin{eqnarray}
&&{\psi}^{(2)}(x,t) = \sqrt{\rho_0} e^{-i\mu_0 \tilde{t}} 
\nonumber \\
&\times& \bigg(1-\frac{G_2(\alpha_0\tilde{t},\sqrt{\alpha_0}x)+i H_2(\alpha_0\tilde{t},\sqrt{\alpha_0}x)}{D_2(\alpha_0\tilde{t},\sqrt{\alpha_0}x)}\bigg),
\label{NLS-second-order}
\end{eqnarray}
where 
\begin{eqnarray}
G_2(x,t)&=&\left(x^2+t^2+\frac{3}{4}\right)\left(x^2+5t^2+\frac{3}{4}\right)-\frac{3}{4},    
\nonumber \\
H_2(x,t)&=&t\left(t^2-3x^2+2(x^2+t^2)^2-\frac{15}{8}\right),
\nonumber \\
D_2(x,t)&=&\frac{1}{3}(x^2+t^2)^3+\frac{1}{4}(x^2-3t^2)^2
\nonumber \\
&+&\frac{3}{64}(12 x^2+44t^2+1),
\nonumber
\end{eqnarray}
are higher-degree polynomial functions.
In particular, Fig.~\ref{fig:1}(d) shows such a comparison revealing minor discrepancies between the two, mainly around the core (as concerns
the width of the first ``blob'') and also
regarding the location of the nodes.

\begin{figure*}
    \centering   \includegraphics[width=\linewidth]{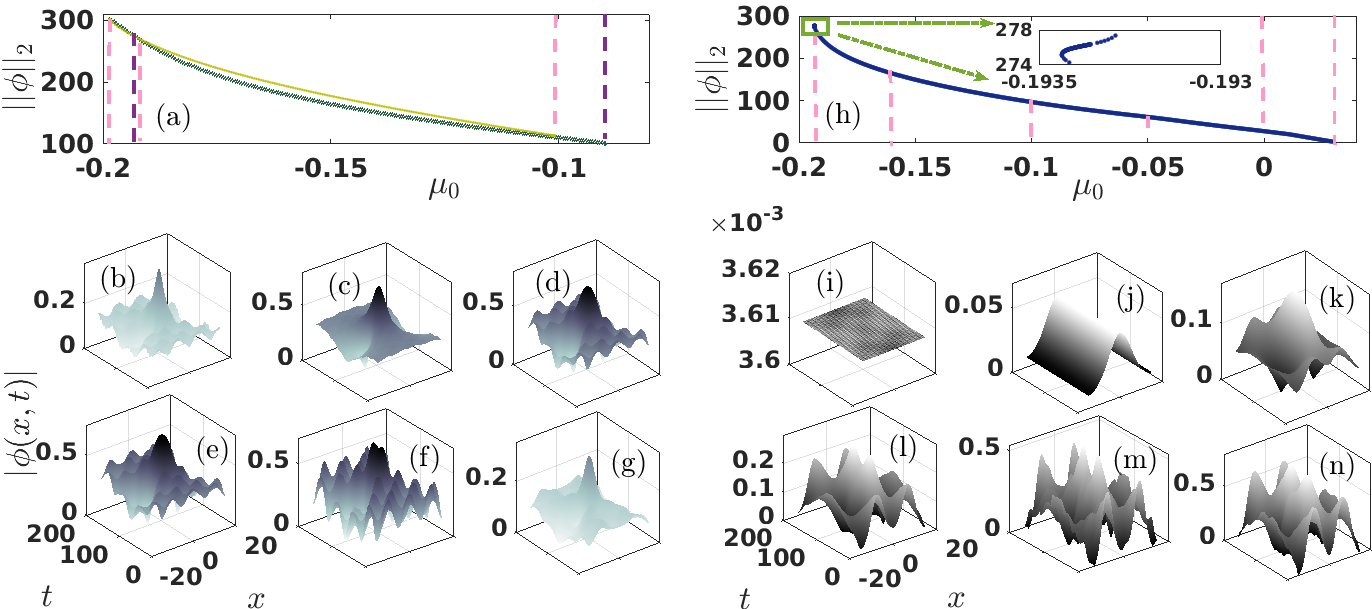}
\caption{{ RW families of the eGPE. The $||\phi||_2$ norm
of the spatiotemporally identified fixed 
point $\phi$ starting from (a) an NLS-like PS (first family) and (h) a sech-like waveform (second family) as a function of the chemical potential $\mu_0$. Inset in (h) shows a magnification of $||\phi||_2$ at $\mu_0\approx-0.1935$, see the green rectangle. Representative waveforms $|\phi(x,t)|$ of the upper (lower) solution branch (i.e., curve) depicted in panel (a) with pink (purple) are shown in panels (b) -(d) ((e)-(g)). 
Here, panel (b) [(d)] depicts a configuration with $\mu_0\approx-0.1009$ [$\mu_0\approx -0.2$] and panel (e) [(g)] refers to $\mu_0\approx-0.2$ [$\mu_0\approx-0.09$].
{We remark that the  purple line (hardly visible) near the turning point at $\mu_0\approx -0.2$ practically coincides  with the pink.} Note the progressive deformation from PS-like structures belonging to the upper branch towards wider RW patterns as $\mu_0$ decreases, and from multi-hump structures (e.g., in (f)) towards PS-like entities (in (g)) in the lower branch as $\mu_0$ increases. In the second RW family, panel (h),  we observe a bifurcation from a homogeneous background in (i) to an NLS periodic wave in (j) and
gradually a PS-like solution for decreasing $\mu_0$ in (k). Multi-hump RWs occur for further decreasing values of $\mu_0$ (panels (l)-(n)). Vertical dashed lines indicate the values of $\mu_0$ for the solutions depicted in panels (b)-(g) and (i)-(n).}}

    \label{fig:red-family}
\end{figure*}
To trace families of RW-like solutions we start from a numerical eGPE RW (converged, upon
iterative usage of the NLS RW approximation)
and subsequently perform parametric continuation over the chemical potential $\mu_0$. Our findings are summarized in Fig.~\ref{fig:red-family}. Specifically, panels (a) and {(h)} within Fig.~\ref{fig:red-family} 
 present the branches of the two distinct families of RW solutions  identified herein. 
An appropriate \textit{matrix norm}, $||\phi||_2=\Bar{\sigma}=\underset{i=1,2,\cdots}{\rm max}[\sigma_{i}]$,where ${\sigma}_i$ represents the singular value decomposition of the discretized $\phi$ (the $L^2$ norm --see, e.g., Ref.~\cite{quarteroni2010numerical}), of each computed solution, is presented against $\mu_0$, yielding a relevant bifurcation
diagram.

Focusing on the first family, see  Fig.~\ref{fig:red-family}(a), the existence curve reveals a simple fold (turning point) near $\mu_0 \approx -0.2$. This turning point signifies the presence of two distinct solution branches, with their corresponding space-time profiles depicted in   Figs.~\ref{fig:red-family}(b)-(d)  and Figs.~\ref{fig:red-family}(e)-(g) respectively. 
The \textit{upper} branch of solutions shown in Figs.~\ref{fig:red-family}(b)–(d) is obtained using the RW breather of Fig.~\ref{fig:1}(b) as an initial guess. Fig.~\ref{fig:red-family}(b) represents the solution for $\mu_0\approx -0.1$, i.e., the solution with the largest chemical potential that was computed on the upper branch. 
As can be seen, the RW solution on the upper branch illustrated in Fig.~\ref{fig:red-family}(b) displays very prominent background modulations. 
On the other hand, as the chemical potential decreases on the upper branch, the amplitude profile of the RW breather broadens, and its peak amplitude also increases, as can be seen in Fig.~\ref{fig:1}(c), where $\mu_0\approx -0.1923$. Eventually, near the turning point $\mu_0\approx -0.2$, the corresponding solution on the upper branch, Fig.~\ref{fig:red-family}(d) has a prominently wide core and rests on a periodic wave background. For such large values of $|\mu_0|$, the core of the eGPE RWs exhibits its most substantial qualitative deviation from their GPE  counterparts.
This is also in line with the reduction of the eGPE to the approximate NLS model described in
Section~\ref{setup}, as the latter is applicable
for smaller amplitudes as in Fig.~\ref{fig:red-family}(b) and becomes progressively less
accurate as the amplitude increases for 
decreasing values of $\mu_0$.

Beyond the turning point, and into the second (lower branch\footnote{Note that past the turning point, there is a small interval in chemical potentials $-0.2<\mu_0<-0.194$ for which the lower branch actually lies above the upper branch.})  $\mu_0\gtrapprox -0.2$, the single-hump structure initially splits along the temporal axis (Fig.~\ref{fig:red-family}(e)), giving rise to a characteristic double-hump profile. Furthermore, this feature appears embedded within a prominent periodic wave background --see Fig.~\ref{fig:red-family}(f), for which $\mu_0\approx-0.1925$. 
{As $|\mu_0|$ is further reduced, the solutions in the lower branch appear to approach GPE-like 
RWs in line with our argument for the
upper branch above; see Fig.~\ref{fig:red-family}(g). These solutions cease to exist at $\mu_0\approx -0.09$ for the chosen parameters.}

The second family of eGPE 
spatio-temporally periodic solutions to which 
our space-time fixed point scheme converges 
is shown in Fig.~\ref{fig:red-family}(h). These bifurcate from the plane waves for $\mu_0\approx 0.03$ (and $\mu_0\ll 1$) and limit to multi-hump periodic waveforms for sufficiently large $|\mu_0|\approx 0.1934$, see Figs.~\ref{fig:red-family}(i)-(n). As the chemical potential crosses from positive to negative values,  
stationary periodic wave solutions are obtained --see Fig.~\ref{fig:red-family}(j). {We have confirmed that these configurations can be well approximated by their cubic GPE counterparts, namely Jacobi elliptic ``dn" functions \cite{chen2020periodic}.} {Furthermore, for negative chemical potentials, small amplitude spatio-temporal solutions are observed, see e.g. Fig.~\ref{fig:red-family}(k) for $\mu_0\approx-0.0505$.} 

For intermediate values of $\mu_0\approx-0.1005$, the central hump splits along the time axis. As such, this doubly periodic waveform {coexists} with a large amplitude periodic wave background --see Fig.~\ref{fig:red-family}(l). With decreasing (more negative) chemical potential, we approach the first turning point ($\mu_0\approx-0.1934$) of the existence curve. The solutions corresponding to these chemical potentials exhibit significant mass concentration in their central humps (Fig.\ref{fig:red-family}(n)). {Extending the continuation beyond the first turning point reveals a second turning point in the bifurcation structure occuring at $\mu_0\approx -0.1933$ (see inset of Fig.~\ref{fig:red-family}(h)). We remark that the qualitative features of the relevant solutions remain unchanged across this second fold (results not shown here for brevity).} 

{Having identified the existence of two distinct families of eGPE RWs, namely the first [second] family shown in Fig~\ref{fig:red-family} (b)-(g) (Fig~\ref{fig:red-family} (i)-(n)), we next aim to inspect their dynamical stability by direct spatio-temporal evolution. Figure~\ref{fig:RW_1_2}(a) presents the evolution of a representative single-hump member (Fig.~\ref{fig:red-family}(c)) of the first family of solutions. As  can be seen, it substantially deforms after two temporal periods. This behavior is a general feature of single hump solutions spanning both families. In contrast, robust propagation is observed (see Fig.~\ref{fig:RW_1_2}(b)) for the two-hump structure depicted in Fig.~\ref{fig:red-family}(l). Evidently, such states survive for several periods before shape deformations manifested in their tails and core take place. Similar dynamics generically pertains to  multihump solutions within both families. More elaborate studies on the (in)stability of these (short-) long-lived (single-) multi-hump solutions in terms of Floquet theory \cite{deconinck_oliveras_2011,karachalios} is deferred to a future work.}

\begin{figure}
\centering
\includegraphics[width=\linewidth]{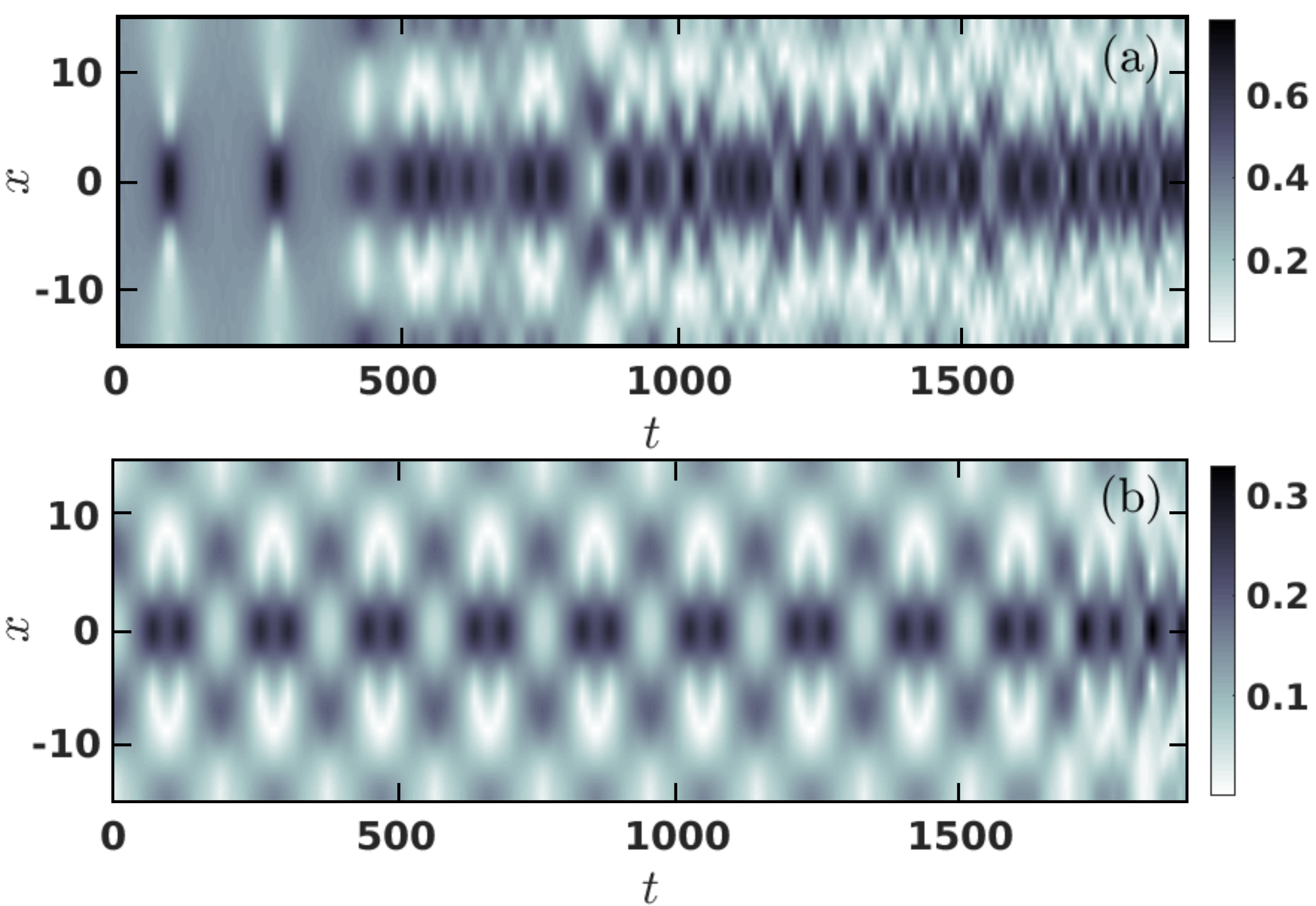}
\caption{{Spatio-temporal evolution of $|\psi(x,t)|$ of (a) a single-humped (Fig.~\ref{fig:red-family}(c)) and (b) a multi-humped (Fig.~\ref{fig:red-family}(l)) RW for $\mu_0 \approx -0.19$ and $\mu_0 \approx -0.1$ respectively. Destabilization of the single-humped RW beyond 2 time-periods is evident, while longevity can be inferred for the multi-hump RW.}}
\label{fig:RW_1_2}
\end{figure}
{The next key step is to explore the dynamical realization of the above discussed RW patterns in physically realistic settings. A persistent challenge here across various attractive environments is the MI of the underlying homogeneous and periodic backgrounds \cite{ward2019evaluating}. Given this, but also
the objective difficulty of directly prescribing
such complex initial conditions, we will now
turn to variants of the dynamical protocols
outlined in the Introduction, to explore 
their potential relevance in the 
competing nonlinearity environment of the eGPE.}
\section{Dynamical RW generation in a droplet environment}\label{dynamics}

Due to the inherent instability of the backgrounds on which rogue-like excitations arise, their dynamical realization remains a significant challenge. 
To address this issue, a recent experimental study exploited a novel dynamical protocol for the production of the PS through a weakly
attractive localized potential in a highly-particle imbalanced repulsive BEC mixture at the immiscibility threshold where effective attractive dynamics is attained~\cite{romero2024experimental}. 
This is in the spirit of
the protocol (i) detailed
in Section~\ref{sec:introduction}, akin
to the proposal of~\cite{ablowitz1990homoclinic}. 

Here, we offer a glimpse at the other
two complementary pathways of Section~\ref{sec:introduction} (i.e., 
protocols (ii) and (iii)) towards dynamically nucleating RWs in droplet bearing eGPE environments. Namely, we pursue the interference of DBFs \cite{el2016dam}, and the onset of a gradient catastrophe \cite{bertola2013universality}. Both of these protocols are based on crafting suitable initial conditions (see below) which can be experimentally realized through the lens of digital micromirror devices~\cite{navon2021quantum,gauthier2016direct}.
Indeed, it has been shown that relevant
protocols leading to the interference
of dam breaks (and to the resulting
formation of PSs~\cite{mossman2024nonlinear})
can be controllably engineered experimentally~\cite{mossman2024nonlinear,tamura2025observationmanybodycoherencequasionedimensional}.

\subsection{Interference of dam breaks}

{DBFs} are modulated periodic wavetrains that arise in attractive media, resembling 
DSWs, typically generated in repulsive interaction regimes \cite{el2016dam, mossman2024nonlinear, wan2010diffraction}. They evolve from an initial discontinuity between a non-zero homogeneous state and a zero background one, and are characterized by a distinct 
``solitonic"-like edge and a linear trailing edge. Despite the MI  
inherent to the attractive regime, these structures persist over moderate time scales \cite{el2016dam,biondini2018riemann,chandramouli2024dispersive}. They are relevant to the universal stage of MI 
in such environments \cite{gurevich1993modulational, biondini2016universal}, and yet their mutual interaction has also been observed to give rise to the formation of a \textit{modulated} RW lattice~\cite{el2016dam}. Leveraging this idea, we construct a uniparametric family of initial ``box" profiles to Eq.~\eqref{GP-eqn} that generates two counterpropagating DBFs,  
\begin{equation}
\label{wavefn-plateuau}
\psi(x,0)=\begin{cases}
\sqrt{\rho_0}, ~~|x|<L\\ 
0,\;\;\;\;\; ~~|x|>L
\end{cases}, 
\end{equation}
where $L=50$ is the considered half box size, unless stated otherwise. 

The extent of the spatial domain ($2L$) determines the characteristic length scale of the developing, modulated RW lattice. 
A key feature of these initial conditions is that the mutual interaction of the emerging DBFs should be 
expected to lead to the formation of RWs, on the basis of the generic nature of the
corresponding integrable NLS results of~\cite{el2016dam}. The box width, $2L$, is selected to be large enough to allow each DBF to develop independently prior to their interaction, yet not so wide that MI effects dominate the interaction stage. The family of initial conditions for the eGPE is chosen to coincide with the interval of existence of DBFs ($0<\rho_0\leq 0.15$). {These bounds are below the hyperbolic-to-elliptic threshold which corresponds to $\rho_0=0.25$ and it is associated with the LHY (attractive) dominating regime according to the recent work of Ref.~\cite{chandramouli2024dispersive}.}

Apparently, this condition sets an upper bound for our dynamical simulations. {Nevertheless, we have confirmed that box-type initial conditions (Eq.~\eqref{wavefn-plateuau}) characterized by $0.15<\rho_0<0.25$ also yield RW configurations. However, these RWs emanate from the interaction of DSW remnant\footnote{{DSW remnants occur across the hyperbolic-to-elliptic threshold encompassing a modulationally stable and an unstable plane wave, see details in \cite{chandramouli2024dispersive}.}} configurations \cite{chandramouli2024dispersive} and, for instance, possess a two-hump spatial structure reminiscent of those illustrated in Fig.~\ref{fig:red-family}(n) instead of the single hump structure occurring from interacting DBFs (see e.g. Fig.~\ref{fig:dam_break1}). This is another fruitful direction to be pursued in future studies, namely the characterization of RW generation through DSW remnant interactions. } 

It turns out that {the} density region {$0<\rho_0\leq 0.15$} can be {divided} further since it is associated with two distinct dynamical regimes.
These correspond to densities 
$\rho_0\ll1/16$ (low density limit) and $\rho_0> 1/16$ (high density limit) respectively, with $\rho_0= 1/16$ designating a threshold above which {DBFs, and thus RWs,} within the droplet setting, are further away from their GPE counterparts (see also Appendix~\ref{Dam-break-flow} for details).

Residing within the low density limit, a representative case example of the interference of the ensuing dam-breaks corresponding to $\rho_0=0.0125\ll1/16$ is illustrated in Fig.~\ref{fig:dam_break1}(a). At the initial stages of the dynamics a modulated pattern is seen (see the relevant profile at $t=128$ in Fig.~\ref{fig:dam_break1}(b)),
in line with the original work
of~\cite{el2016dam} and the recent
realizations of~\cite{mossman2024nonlinear,tamura2025observationmanybodycoherencequasionedimensional}.
Within this modulated waveform a central peak entity can be discerned. 
The latter bears a {RW}-like signature,
clearly evident when compared to the numerically obtained configuration for $\mu_0 \approx -0.0689$. {Here the corresponding eGPE RW was obtained as a fixed point to Eq.~\eqref{eGPE-breather} using computational parameters $L=40$ and $T=190$}. 

{At later stages, during the spatiotemporal evolution, the pattern evolves into a {modulated} RW lattice, in close analogy to its GPE counterpart~\cite{el2016dam}. This becomes evident by comparing the dynamical profile at $t=244$ to a snapshot of a computed eGPE RW lattice at $\mu_0\approx -0.085$ where good agreement between the two is observed (Fig.~\ref{fig:dam_break1}(c)). We remark that for this boundary value problem computation, an Akhmediev breather (AB) waveform was used 
\begin{align}
\label{AB-NLS}
&\psi^{(\rm AB)}(x,t)\\\nonumber &=\sqrt{\rho_0}\left(1+\frac{2(1-2a)+ib\tanh(b\alpha_0\tilde{t})}{\sqrt{2a}{\rm sech}(b\alpha_0\tilde{t})\cos(\omega\sqrt{\alpha_0} x)-1}\right)e^{-i\mu_0 \tilde{t}},
\end{align}
as an initial guess together with the computational parameters $L\approx92.69$ and $T=200$. Here, the AB breather is characterized by two free parameters. Namely, the background density $\rho_0$ and the control parameter\footnote{Tuning $a$ towards $0.5$ leads to the PS. } $0<a<0.5$, while $\omega=2\sqrt{1-2a}$ and $b=\sqrt{8a(1-2a)}$ 
 are determined by $a$. Note also the excellent agreement between the central lattice element (i.e., centered about $x=0$) and a numerically identified PS-like wave obtained from the boundary value problem  with $\mu_0\approx -0.0767$, $L=40$, and $T=190$ (Fig.~\ref{fig:dam_break1}(c)). This observation signals the progressive deformation and in particular, the lengthening of the spatial period of the \textit{modulated} RW lattice \cite{el2016dam}.}

Additionally, the dynamical formation of the RW lattice configuration is accompanied by the emission of counterpropagating matter wave jets. These are a manifestation of the self-evaporation~\cite{ferioli2020dynamical} of the droplet background due to its excited nature induced by the interference process. Such self-evaporation mechanism, leading
to density emission from a 1D droplet core has also been argued to occur after droplet collision~\cite{astrakharchik2018dynamics,katsimiga2023interactions} and in the presence of a dynamically switched-on defect~\cite{bristy2025localization}. 
It is worth mentioning at this point that an overall similar response takes place within the standard GPE model (not shown) but without the above-discussed emission of droplet-like density portions. 
This further supports the importance of the LHY contribution and marks the
differences of the phenomenology in the
presence of competing nonlinear interactions. 
\begin{figure}
\centering    \includegraphics[width=\linewidth]{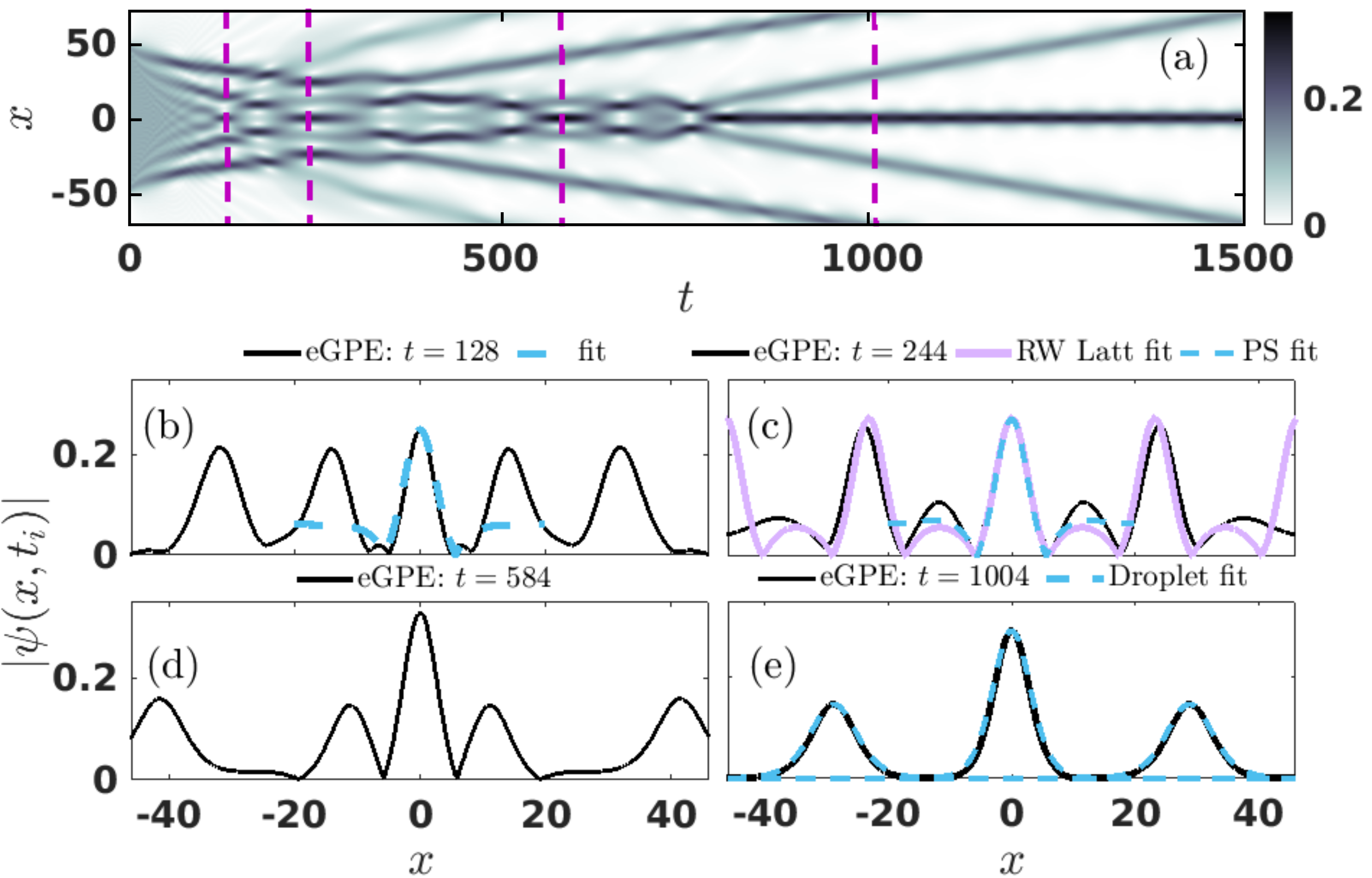}
\caption{{(a) Interfering DBFs leading to modulated RW lattice and PS generation for $\rho_0=0.0125$. 
(b) Snapshot of $|\psi|$ revealing the formation of an eGPE RW compared to a numerically obtained PS-like wave with $\mu_0\approx -0.0689$. (c) Relevant profile demonstrating a RW lattice {compared to a computed eGPE one having $\mu_0\approx -0.085$. The central breather is also compared to a computed eGPE PS at $\mu_0\approx -0.0767$. (d) A 
structure reminiscent of a higher order solitonic waveform, and (e) a temporally modulated, persistent quantum droplet appears at the center,
accompanied by symmetrically emitted matter wave jets, with the central and the lateral structures fitted to the  analytical droplet profile with $\mu_0\approx-0.1518$ and $\mu_0\approx -0.087$ respectively}. Vertical dashed lines in panel (a) indicate the times used for the profiles shown in panels (b)-(e).}}
    \label{fig:dam_break1}
\end{figure}

As time progresses, instead of a recurrence of the PS-like RW, a higher order solitonic
structure appears to be formed around $t=584$ [Fig.~\ref{fig:dam_break1}(d)] that re-emerges at $t = 808$ along with the new emission of counterpropagating matter wave jets [see Fig.~\ref{fig:dam_break1}(a)], leaving behind a central breathing configuration. 
We were able to identify the nature of this central configuration to a {breathing} droplet that persists throughout the evolution, {while on the other hand the counterpropagating jets were identified as \textit{traveling-breathing} ones } [see Fig.~\ref{fig:dam_break1}(e)].} To support our arguments in Fig.~\ref{fig:dam_break1}(e), the stationary droplet profile~\cite{astrakharchik2018dynamics} 
\begin{equation}
|\psi|=
\frac{3\mu_0}{1+\sqrt{1+\frac{9\mu_0}{2}}\cosh(\sqrt{-2 \mu_0}x)},
\end{equation}
with $\mu_0 \approx -0.1518$ is overlaid atop the dynamically formed breather evincing excellent agreement between the two. 
\begin{figure}
\centering    \includegraphics[width=\linewidth]{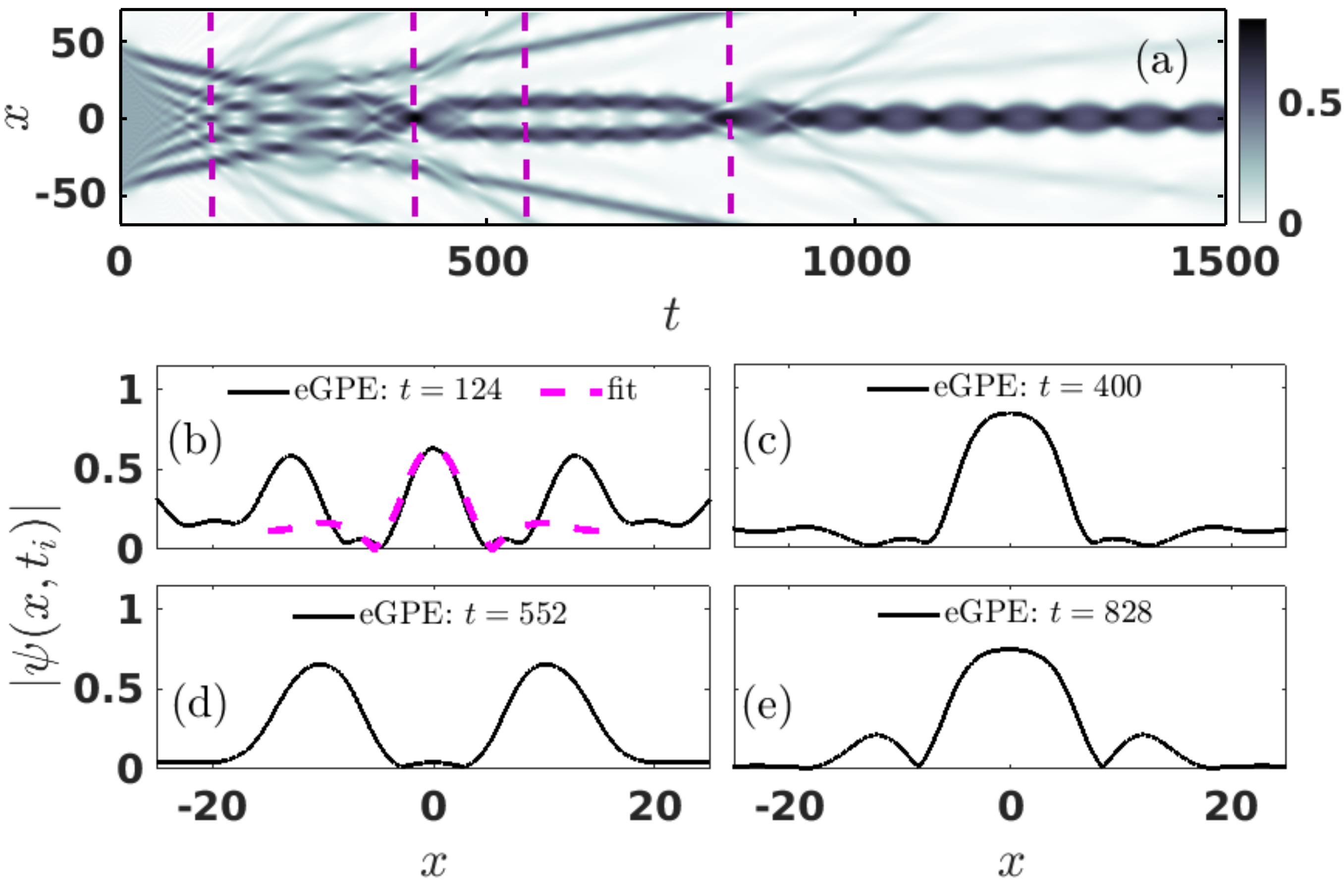}
\caption{{(a) Interfering DBFs for $\rho_0=0.0875$ leading to RW  generation.
(b) Snapshot of $|\psi|$ featuring an eGPE RW which is compared with the computed rogue waveform shown in Fig.~\ref{fig:red-family}(a) having $\mu_0\approx -0.1713$. (c) A structure
reminiscent of a HORW. (d) The snapshot illustrates a bimodal breathing state, which subsequently merges into (e) a larger amplitude structure
centered at $x=0$. Vertical dashed lines in panel (a) mark the time-instants presented in panels (b)-(e).}}
    \label{fig:interfering_dam_2}
\end{figure}

Turning to the high-density limit the dynamical response entails the nontrivial modification of the aforementioned waveforms. 
A prototypical example of the ensuing counterflow process is demonstrated in Fig.~\ref{fig:interfering_dam_2}(a) for
$\rho_0=0.0875>1/16$.
Notably, as captured by the snapshot shown at $t=124$ in
Fig.~\ref{fig:interfering_dam_2}(b)
a significantly wider RW is generated. This core alteration can be attributed to the competition between mean-field repulsion and LHY attraction that becomes more impactful in this higher-density limit.
This broad RW is also compared to
the numerically obtained solution
corresponding to $\mu_0\approx -0.1713$ and illustrated in the same figure by a dashed line demonstrating very good agreement. As in the low density scenario, the development of a modulated rogue lattice consisting of broad RW structures 
is observed within $120\leq t <400$. Also here, the process is accompanied by matter wave jets caused by self-evaporation.

Finally, around $t=400$, a 
structure with multiple nodes, 
somewhat reminiscent
of a HORW appears; see Fig.~\ref{fig:interfering_dam_2}(c). For even longer times depicted in Fig.~\ref{fig:interfering_dam_2}(d), a two hump configuration 
is evidenced merging around $t=828$ toward a larger
amplitude blob centered at $x=0$
[Fig.~\ref{fig:interfering_dam_2}(e)] whose core periodically expands and contracts  being persistent throughout the evolution, see Fig.~\ref{fig:interfering_dam_2}(a), and in particular for all times up to $t=5000$ that we have checked. 
Indeed, this state bears the features
of a two-soliton breathing waveform~\cite{satsuma}. The generation of related localized breathing
waveforms has been reported in earlier works 
within the eGPE model~\cite{astrakharchik2018dynamics,katsimiga2023interactions}; yet, to the best of our knowledge, they have not been identified as part of a family of coherent excitations, nor have their stability properties been investigated.
This is an intriguing possibility for future
studies in its own right.

\subsection{Gradient Catastrophe}

Next, we explore yet another way to dynamically generate RWs. This is done by examining the long-time evolution of a monoparametric family of initial conditions having the form:
\begin{equation}
\label{Sech-waveform}
    \psi(x,0)=A {\rm sech}(x),
\end{equation}
where $A$ denotes the   
amplitude of the localized waveform. 
We note in passing that in repulsive environments, localized waveforms are known to broaden \cite{ivanov2019collision}, leading to the long-time decay of the peak amplitude (since $\int_{\mathbb{R}} |\psi|^2 dx$ is a conserved quantity).  {In contrast, in attractive media governed by the cubic focusing GPE, the interplay between nonlinear self-focusing —which tends to steepen the waveform— and quantum pressure-induced dispersion leads to the formation of a RW structure \cite{tikan2017universality,bertola2013universality,el2016dam,demiquel2025gradient}. At later times, this structure evolves into an increasingly intricate 
RW lattice \cite{bertola2013universality,el2016dam}.}

At the outset, we note that the peak amplitude $A$ of the initial waveform is of particular interest, as its value can determine whether the repulsive mean-field term or the attractive LHY term dominates the \textit{short-time} dynamics. In this work, we focus on the regime where $A>0.5$, which lies above the hyperbolic-elliptic threshold \cite{chandramouli2024dispersive}, {meaning that mean-field repulsion dominates}. {However, as we will show, under this condition, repulsive wavepacket behavior is observed over short time scales, transitioning to effectively attractive dynamics at longer times, facilitating RW generation.} {On the other hand, for $A<0.5$, LHY attraction naturally  prevails, leading to RW formation (not shown).} 

\begin{figure}
\centering
\includegraphics[width=\linewidth]{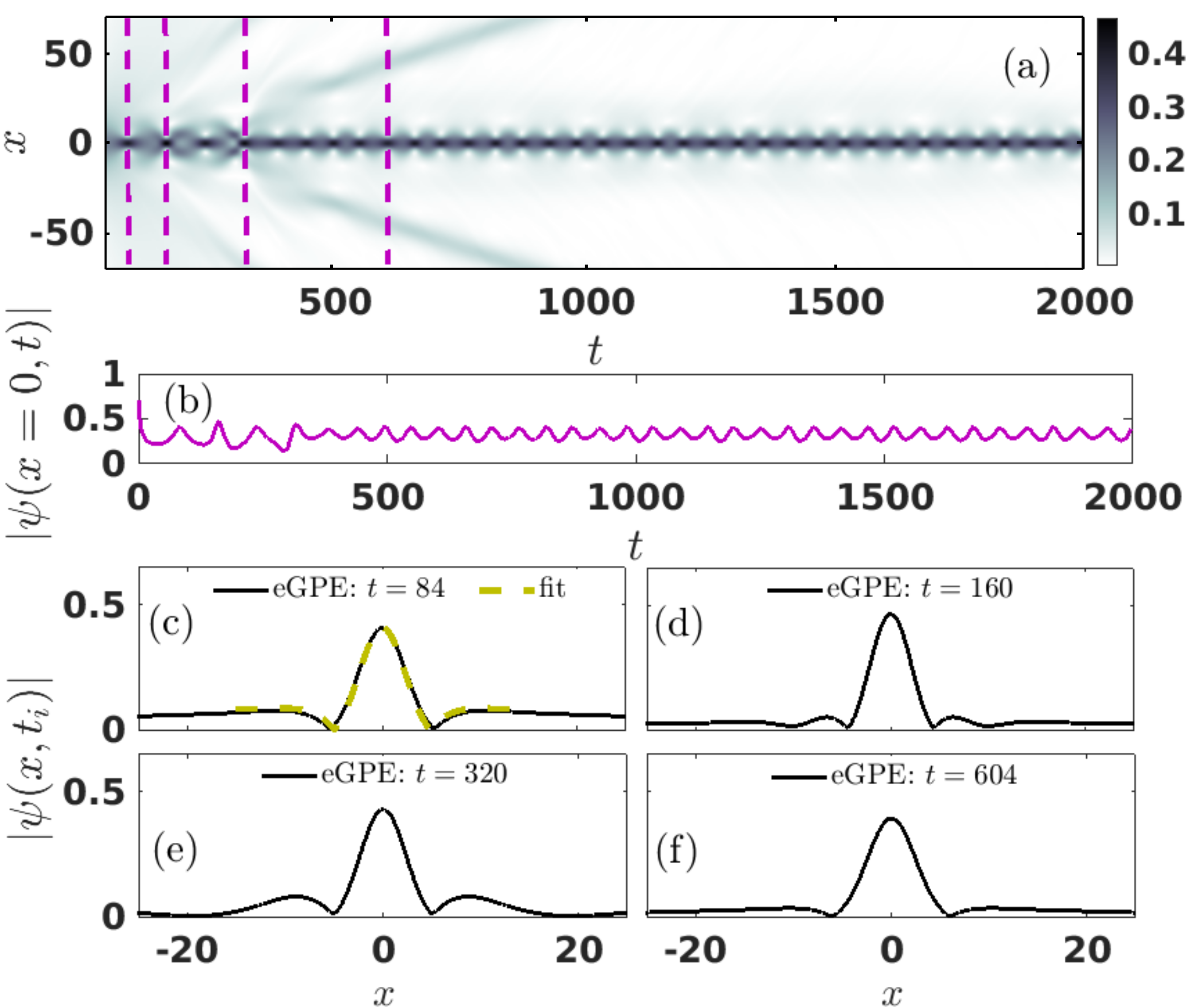}
\caption{{Gradient catastrophe of a localized wavepacket (Eq.~\eqref{Sech-waveform}) leading to PS and HORW generation. (a) Spatiotemporal evolution of $|\psi|$. (b) Dynamics of $|\psi(x=0,t)|$ dictating repulsion (mean-field term) over short time-scales followed by attraction (driven by the LHY term) 
as evidenced by the quasi-periodic recurrence. (c) Snapshot depicting the formation of an eGPE PS due to the gradient catastrophe mechanism; and compared to an eGPE fixed point solution (c.f. Fig.~\ref{fig:red-family} (a)), with $\mu_0=-0.1138$. (d) HORW profile, whose recurrence is shown in panel (e). 
(f) In the long-time dynamics equilibration is observed toward a time-periodic uninodal waveform. Vertical dashed lines in panel (a) mark the time-instants presented in panels (c)-(f).}}
    \label{fig:grad_catst}
\end{figure}
Figure~\ref{fig:grad_catst}(a) shows the spatiotemporal evolution (for $t\geq 40$) of the wavefuction amplitude, $|\psi(x,t)|$, when Eq.~(\ref{Sech-waveform}) is utilized as an initial ansatz with $A=1/\sqrt{2}$. {Here, the initial, repulsive dynamical response for $t<40$ has been omitted to emphasize the effectively attractive dynamics thereafter.} However, the repulsive behavior at short time scales {can be seen in the initial decay of the wavefunction peak amplitude in Fig.~\ref{fig:grad_catst}(b).} For $t>40$, the LHY-term dictates the dynamics, leading to a quasi-periodic recurrence evident in Fig.~\ref{fig:grad_catst}(a) and in the central peak wavefunction amplitude depicted in Fig.~\ref{fig:grad_catst}(b).

The underlying gradient catastrophe mechanism {within this} attractive regime leads to the formation of a rogue event, at $t = 84$ (Fig.~\ref{fig:grad_catst} (c)). This is compared to a numerically computed eGPE RW, shown as a yellow dashed line, obtained via a fixed point computation with $\mu_0 = -0.1138$ (see Fig.~\ref{fig:red-family}(a)), demonstrating 
very good agreement. At $t = 160$, an event reminiscent of a HORW emerges, which then recurs ---with the
nodal points expanding outwards--- at $t = 320$ (Figs.~\ref{fig:grad_catst}(d),(e)). 
Over longer times, a highly coherent nearly time-periodic excitation forms as can be seen by the wavefunction amplitude snapshot presented in Fig.~\ref{fig:grad_catst}(f); this 
is again reminiscent of two-soliton breathing
solutions within {the integrable, cubic} NLS~\cite{satsuma}.

{Concluding, our studies have focused on the case of unit width of the initial wavepacket [Eq.~\eqref{Sech-waveform}]. This choice places us away from the semi-classical limit investigated in \cite{bertola2013universality}. 
Nevertheless, this model possesses interesting dynamics because
the defocusing initial dynamics of the wavepacket spreads it while
bringing its amplitude within the realm of the focusing-dominated
range of amplitudes. This, in turn, leads to the observed 
gradient catastrophe dynamics observed above.
Certainly a detailed exploration of the impact of the width parameter is required in the future to infer, for instance, the potential presence of a ``Christmas-tree" structure \cite{bertola2013universality,tikan2021local,el2016dam,charalampidis2016rogue}. 
}

\section{Conclusions and  future perspectives}\label{conclusions}

We have investigated the emergence of the (so far) elusive RWs in a 1D symmetric eGPE model
describing a
quantum droplet bearing environment. The mathematical setup has the form of a  NLS/GPE model containing competing repulsive mean-field and attractive LHY nonlinear terms. 
The physical setting of interest is associated with symmetric homonuclear bosonic mixtures bearing equal populations and masses per component as well as the same intracomponent repulsive interaction strengths. This ensures the modeling of the system by a reduced single-component eGPE. 

Specifically, we have computed families of RWs using a fixed-point scheme (in space and time,
considering time as a spatial dimension) 
based on a doubly periodic basis representation. 
Upon chemical potential variations, distinct regions, in particular $|\mu_0|\ll 1$, are identified where the ensuing RWs resemble the standard PS of the cubic NLS/GPE model. 
Importantly, we explicate that there are chemical potential intervals, $\mu_0\gtrsim -0.12$, in which the identified RWs {resemble very closely} their NLS/GPE counterparts.   
These findings have been corroborated by the \textit{effective} cubic GPE  approximation obtained out of the utilized eGPE model. 
Unlike the cubic GPE, RWs in the droplet environment arise under a non-trivial influence of the repulsive mean-field term, and thus are wider than their cubic GPE siblings. 

To demonstrate possible dynamical protocols facilitating the experimental observation of these extreme events, we next simulated: i) two-counterpropagating DBFs with Riemann initial conditions and ii) the gradient catastrophe of highly localized waveforms. 
Such initial conditions can be ---
and in some cases have been~\cite{mossman2024nonlinear,tamura2025observationmanybodycoherencequasionedimensional}--- 
readily sculpted with the aid of modern ultracold atom platforms. 
Crucially, both of these do not rely on the MI assisted RW production of homogeneous backgrounds and hence avoid possible experimental complications that accompany this type
of initialization. 
Both mechanisms lead to a variety of RW excitations, including modulated rogue lattices (reminiscent of the Akhmediev-breathers of the cubic GPE), second-order RWs to the eGPE, and localized periodic 
waveforms (reminiscent of multisolitons) persisting for long evolution times. 
Despite the shared phenomenology of the mechanisms of RW generation, there are various non-trivial differences in the dynamics with the \textit{semiclassical} cubic GPE \cite{bertola2013universality,el2016dam}, due to the difference in the operating regime, the lack of \textit{integrability} of the  eGPE and the influence of the mean-field repulsion. 

There are several interesting future research pathways based on our findings which only set the stage for RW formation in systems with competing interactions. 
A straightforward relevant direction is to adapt a doubly rational basis representation, extending the work of Ref.~\cite{cole2024approximation}, along with a fixed point scheme in order to achieve increased accuracy of the rogue waveforms, particularly near the tails where there is an anticipated algebraic decay to the background wave. This could also enable a continuation
of the obtained families 
past the termination points found herein.
Such a study could be corroborated by variational methods, similarly to Ref.~\cite{bokaeeyan2019rogue}, yielding potential insights into parametric regimes of existence of RWs. 
Yet another avenue of immediate relevance is to \textit{systematically} compute families of RWs coexisting with eGPE periodic waves (similar to cnoidal RWs in the cubic GPE~\cite{chen2018rogue,chen2019rogue}). 
In this setting, it may be relevant 
to identify the doubly periodic eGPE solutions
first, similarly, e.g., to 
the work of~\cite{ward2020rogue}.
A related and important direction for further study is the RW formation by means of the MI in droplet-bearing environments, extending earlier work of Refs.~\cite{mithun2020modulational,mithun2021statistical}. 
In particular, it would be of interest to investigate the growth of secondary waves using an asymptotically valid reduction (a complex Ginzburg-Landau equation) for the background envelope~\cite{aranson2002world}.

Higher-dimensional variants of the eGPE \cite{petrov2016ultradilute,petrov2015quantum} exhibit broader parameter regimes supporting attractive interactions and may provide a fertile ground for the emergence of rogue waveforms. 
Here, it is relevant to examine the phenomenology of {\it quasi-one-dimensional solutions}
that are uniform in the transverse direction,
as well as of genuinely two-dimensional 
solutions.
In particular, it remains an open question whether interfering (radial) dam-break dynamics could be harnessed to facilitate dynamical nucleation of RWs. 
Finally, applying nonperturbative ab-initio  approaches~\cite{mistakidis2021formation,englezos2023correlated} to investigate the role of beyond-LHY correlations in the dynamical nucleation of RWs e.g., via the interference of quantum dam breaks is a highly compelling direction which is currently lacking in the literature.   
Such studies are currently in progress
and will be reported in future publications.

\appendix

\section{Dam breaks in the attractive regime} \label{appendix}
\begin{figure}
    \centering
\includegraphics[width=\linewidth]{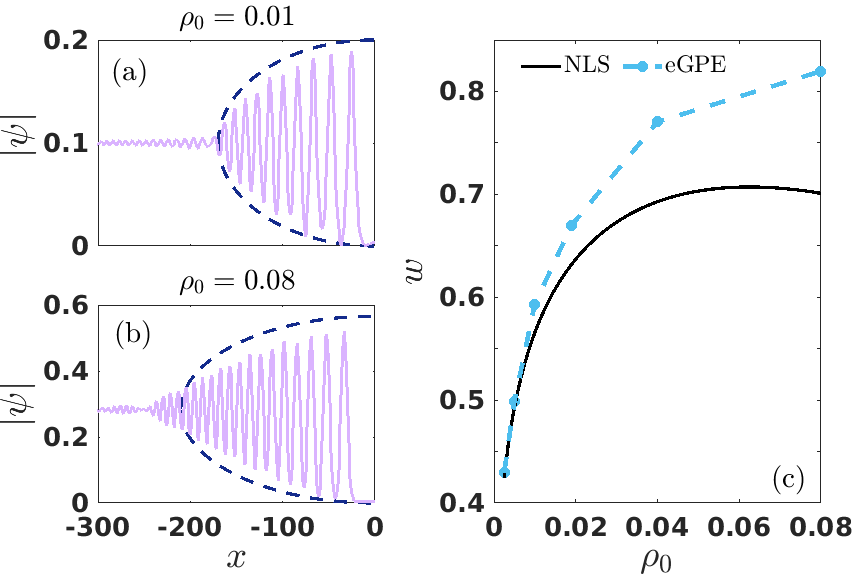}
\caption{{Moduli, $|\psi|$, snapshots of distinct types of DBFs governed by the 
eGPE utilizing a Riemann initial condition ($\rho_0$,0).
(a) A representative low-density case ($\rho_0=0.01$) showing the eGPE DBF at $t=300$. Good agreement with the approximate NLS 
envelope [Eq.~(\ref{Oscillation-envelope})]—both in shape and in  the spreading rate  [Eq.~(\ref{Eq:vel_width})] occurs. (b) A characteristic  intermediate-density case ($\rho_0=0.08$) at $t=300$.  Noticeable discrepancies emerge between the eGPE and the NLS envelopes. (c) {The spreading rate},  $w=v_+-v_-$, as a function of the  initial density $\rho_0$, comparing the approximate NLS [Eq.~(\ref{cubic-GP})] prediction of  Eq.~\eqref{Eq:vel_width}(solid curve) with the eGPE outcome (dashed dotted line). A clear deviation from the analytical result manifests as $\rho_0$ increases.}  }
    \label{fig:dam-break-flow-eGPE-comparison-to-GP}
\end{figure}
\label{Dam-break-flow}
{
The cubic focusing NLS 
\begin{equation}
    i\psi_{t'}+\frac{1}{2}\psi_{x'x'}+|\psi|^2\psi=0,
\end{equation}
hosts DBFs, namely modulated periodic waves with distinct linear and bright solitonic fronts. This self-similar modulation solution is defined implicitly through the slowly varying elliptic modulus $m$ (c.f. \cite{el2016dam}) 
\begin{align}
 \frac{x'}{t'}&=-\frac{2\sqrt{\rho_0}}{m\mu(m)}\sqrt{(1-m)(\mu^2(m)+m-1)}\times \\\nonumber &\left(1+\frac{(2-m)\mu(m)-2(1-m)}{\mu^2(m)+m-1}\right),
\end{align}
where $K(m)$ and $E(m)$ represent the complete elliptic integrals of the first and second kind, respectively, while $\mu=\frac{E(m)}{K(m)}$. Having ascertained the slowly varying elliptic modulus parameter, one can also obtain the oscillation (upper and lower) envelopes \cite{gurevich1993modulational,el2016dam} via
\begin{align}
\label{Oscillation-envelope}
    |q_{\pm}(m)|=\sqrt{\rho_0}\pm \frac{\sqrt{\rho_0}}{m}\left[2-m-2(1-m)\frac{K(m)}{E(m)}\right].
\end{align}
Rescaling, the spatial coordinate in the approximate NLS of Eq.~\eqref{cubic-GP} as $x=\sqrt{f'(\rho_0)} x'$, it is possible to extract the envelopes of the DBFs in this rescaled coordinate system. In particular, the {spreading rate, defined as the } difference in edge speeds 
between leading and trailing edge is given by  
\begin{equation}\label{Eq:vel_width}
w= v_+-v_-=2\sqrt{2\rho_0 f'(\rho_0)},
\end{equation}
 where $v_+$ ($v_-$) are the solitonic (linear) edge speeds of the DBF.} {Notably, this approximate spreading rate curve has a maximum at $\rho_0=1/16$, beyond which the qualitative (and quantitative) behavior differs significantly from the numerically extracted eGPE analog.} 
 
 {To elaborate on the qualitatively distinct density regimes, we show representative DBFs in Fig.~\ref{fig:dam-break-flow-eGPE-comparison-to-GP}(a) ($\rho_0 = 0.01 \ll 1/16$) and Fig.~\ref{fig:dam-break-flow-eGPE-comparison-to-GP}(b) ($\rho_0 = 0.08 > 1/16$), respectively. In the low-density limit, the envelopes and spreading rates  show good agreement with the predictions of the approximate cubic NLS model given by the rescaled Eq.~\eqref{Oscillation-envelope}, see Fig.~\ref{fig:dam-break-flow-eGPE-comparison-to-GP}(a). In contrast, for $\rho_0 > 1/16$, there is a clear mismatch in the envelope shapes but a less pronounced discrepancy in the spreading rate (Fig.~\ref{fig:dam-break-flow-eGPE-comparison-to-GP}(b)).} {A deeper understanding of these discrepancies is desirable, calling upon further studies utilizing the relevant Whitham modulation equations in the attractive regime~\cite{el2016dam,gurevich1993modulational}.}

{To provide an overview of the applicability of the NLS picture (Fig.~\ref{fig:dam-break-flow-eGPE-comparison-to-GP}(c)), in the considered parametric regime, we next examine the variation in the spreading rate of the DBF's envelopes as a function of $\rho_0$. Evidently, excellent agreement between the approximate NLS prediction and the eGPE outcome is observed within the low-density limit. Furthermore, for intermediate densities $0.01\lessapprox\rho_0<1/16$, the two spreading rate curves follow a similar increasing trend, thus demonstrating qualitative agreement. Beyond this point, deviations between the spreading rates in the two models become more prominent. Specifically, the NLS predicts a decreasing spreading rate in accordance with the rescaling $x=\sqrt{f'(\rho_0)}x'$, in contrast to the progressively increasing one within the eGPE.}

\section{DATA AVAILABILITY}
The data that support the findings of this article are not publicly available upon publication because it is not technically feasible and/or the cost of preparing, depositing, and hosting the data would be prohibitive within the terms of this
research project. The data are available from the authors upon
reasonable request.

\acknowledgements
The authors are thankful to M. A. Hoefer and E. G. Charalampidis for insightful discussions. {This material is based upon work supported by the US National Science Foundation under Grant No. PHY-
2110030, PHY-2408988 and DMS-2204702 (P.G.K.).}
S.I.M. acknowledges support from the Missouri University of Science and Technology, Department of Physics, Startup fund. This research was partly conducted while P.G.K. was 
visiting the Okinawa Institute of Science and Technology (OIST) through the Theoretical Sciences Visiting Program (TSVP). This work was also supported by the Simons Foundation [SFI-MPS-SFM-00011048, P.G.K.].

\bibliography{SHOCK_WAVES_QUANTUM_DROPLET}
\end{document}